\documentclass[journal=jctc,manuscript=article]{achemso}

\usepackage[version=3]{mhchem} % Formula subscripts using \ce{}
\usepackage{physics}           % Bra ket notation

\newcommand{\T}{\hat{T}}
\newcommand{\rf}{\ket{\Phi_0}}
\newcommand{\kcal}{kcal mol$^{-1}$}
\newcommand{\cm}{cm$^{-1}$}
\newcommand{\Eh}{$mE_\mathrm{h}$}

\author{Gustavo J.R. Aroeira}
\author{Madeline M. Davis}
\author{Justin M. Turney}
\author{Henry F. Schaefer III}
\email{ccq@uga.edu}
\phone{+1 706 542 0406}
\fax{+1 706 542 2067}
\affiliation[University of Georgia]
{Center for Computational Quantum Chemistry, University of Georgia, Athens}

\title{Coupled Cluster Externally Corrected by Adaptive Configuration Interaction}

\abbreviations{IR,NMR,UV}

\begin{document}
%%%%%%%%%%%%%%%%%%%%%%%%%%%%%%%%%%%%%%%%%%%%%%%%%%%%%%%%%%%%%%%%%%%%%
%% The manuscript does not need to include \maketitle, which is
%% executed automatically.  The document should begin with an
%% abstract, if appropriate.  If one is given and should not be, the
%% contents will be gobbled.
%%%%%%%%%%%%%%%%%%%%%%%%%%%%%%%%%%%%%%%%%%%%%%%%%%%%%%%%%%%%%%%%%%%%%
\begin{abstract}
An externally corrected coupled cluster (CC) method, where an adaptive configuration interaction (ACI) wave function provides the external cluster amplitudes, named ACI-CC, is presented. By exploiting the connection between configuration interaction and coupled cluster through cluster analysis, the higher-order $T_3$ and $T_4$ terms obtained from ACI are used to augment the $T_1$ and $T_2$ amplitude equations from traditional coupled cluster. These higher-order contributions are kept frozen during the coupled cluster iterations and do not contribute to an increased cost with respect to CCSD. We have benchmarked this method on three closed-shell systems: beryllium dimer, carbonyl oxide, and cyclobutadiene, with good results compared to other corrected coupled cluster methods. In all cases, the inclusion of these external corrections improved upon the “gold standard” CCSD(T) results, indicating that ACI-CCSD(T) can be used to assess strong correlation effects in a system and as an inexpensive starting point for more complex external corrections.
\end{abstract}

\section{Introduction}
Since its introduction in quantum chemistry, coupled-cluster theory (CC)\cite{cizek1966,cizek1969,cizek1971,Paldus1972} has become a method routinely employed to compute reliable correlation energies. The single reference (SR) coupled-cluster including single and double excitations (CCSD)\cite{Purvis1982,Scuseria1987,Scuseria1988} method can provide satisfactory results for various molecular properties. These results are often superior to those from alternative methods, such as configuration interaction with single and double excitations (CISD), or second-order M{\o}ller--Plesset perturbation theory (MP2).\cite{Bartlett2007} The contribution of high-order excitations can be included in an economical way using energy correcting methods such as the CCSD with perturbative triples [CCSD(T)] approach.\cite{Raghavachari1989,Bartlett1990} This method has become a popular option to compute accurate energies, in particular for nondegenerate closed-shell systems near their equilibrium geometries. Thus, the CCSD(T) method is often regarded as the ``gold standard"of quantum chemistry.\cite{Gauss1998} However, many properties require energies at different points on the potential energy surface (PES). Despite the effectiveness of the CCSD(T) at equilibrium geometries, the method may break down, for instance, when bonds are significantly stretched.\cite{Ghose1995} For so-called multireference (MR) or strongly correlated systems, such as the bond-breaking scenario, the reference determinant loses its dominance. As a consequence, excitations beyond double become increasingly important and the inclusion of high-order contributions via perturbative methods becomes ineffective. 

In order to improve the description of strongly correlated systems while maintaining the SR framework one can explicitly account for high-order excitations. Methods such as CCSDT,\cite{Scuseria1988b,Noga1987,Watts1990} CCSDTQ,\cite{Oliphant1991,Kucharski1991,Kucharski1992} CCSDTQP,\cite{Musia2002} and higher can successfully solve many multireference problems, such as quasidegeneracy. Unfortunately, these approaches are computationally very demanding and, generally, can only be used with relatively small basis sets. Alternative approaches, such as CCSDt and CCSDtq,\cite{Piecuch1999} aim to reduce this cost by defining active spaces for the higher-order cluster operators. For example, the CCSDt method uses the same equations as CCSDT, but include only a small subset of the full $\T_3$ amplitudes (thus ``t'' instead of ``T''). However, even with this augmentation, the description of systems like \ce{N2} with triple bonds can still be less than desirable. To properly account for these quasidegeneracies, a MR CC formulation is needed. Efforts to formulate such methods have given rise to a family of MR CC approaches each with different achievements and shortcomings.\cite{Bartlett2007,Lyakh2012} 

Externally corrected coupled-cluster (ecCC) has been proposed as a strategy to recover higher-order excitation contributions to the CC energy. The premise is to use a separate method from which  three- and four-body cluster amplitudes can be extracted and introduced into the CC equations. An early application of this strategy used unrestricted Hartree--Fock wave functions to estimate quadruple-excitation contributions to the coupled-pair many-electron theory.\cite{Paldus1984} Later, Paldus and Planelles employed valence bond wave functions as the source of the higher-order amplitudes.\cite{Paldus1994,Planelles1994a,Planelles1994b} The usage of complete active space self-consistent field (CASSCF) as the source for the external amplitudes was suggested by Stolarczyk\cite{Stolarczyk1994} in his proposal of a complete active space coupled-cluster (CASCC) theory. However, this strategy was only implemented and explored by Peris and coworkers, two years later.\cite{Peris1997} This idea was further pursued by Li and Paldus\cite{Li1997} in their reduced multireference (RMR) CC method, which can be described as a step-by-step procedure: (a) a multireference configuration interaction (MRCI) wave function is constructed within an active space; (b) using cluster analysis,  $T_3$ and $T_4$ amplitudes are extracted; (c) a CCSD computation is performed, including the three- and four-body contributions generated previously, in a CCSDTQ-like framework. This approach was shown to outperform both MRCI and CCSD in several systems,\cite{Li1997,Li1998,Li2003,Li2007} including a very accurate description of the \ce{N2} bond breaking scenario.\cite{Li2008} Further examples and discussion on the RMR CC method can be found in literature reviews.\cite{Paldus1999,Paldus2010} Peris \textit{et al.}\cite{Peris1999} also proposed the use of perturbative selected configuration interaction (CIPSI)\cite{Huron1973} to reduced the dimension of the CI wave function. Their results for model systems were within 1 \Eh{} of the RMR CCSD methods while considerably reducing the cost of the computation.  Xu and Li\cite{Xu2015} demonstrated that the ecCC approach can be extended to even higher-order external amplitudes. They used CASSCF to generate $\T_4$ and $\T_5$ terms that were introduced into a CCSDt routine yielding results in good agreement with experiment for a selected set of strongly correlated systems.

The split amplitude strategy of the ecCC methods was implemented in a different way by Kinoshita and coworkers.\cite{Kinoshita2005} In their tailored coupled-cluster method (TCC), a complete active space configuration interaction (CASCI) computation generates the external correction. However, unlike the RMR CC method, only $T_1$ and $T_2$ amplitudes are extracted from the CASCI computation. These amplitudes are then constrained to be unchanged during the CC iterations. The method relies on the assumption that the CAS amplitudes ($T_1$ and $T_2$) will retain their multireference information and will continue to describe the strong correlation aspects of the system. The remaining $T_1$ and $T_2$ amplitudes (outside the CAS space) are then used to obtain the dynamical component of the correlation energy. Despite its simplicity, potential energy surfaces and dissociation energies obtained with this method are in good agreement with the more demanding MRCI approach.\cite{Kinoshita2005} However, due to the decoupling of the two set of amplitudes,  TCC presents a large nonparallelity error (NPE).\cite{Melnichuk2012}

More recently, Veis and coworkers\cite{Veis2016} combined the  TCC split amplitude strategy with density matrix renormalization group (DMRG). This version of TCC uses DMRG as the source of $T_1$ and $T_2$ amplitudes, this approach has the advantage that DMRG can handle large active spaces with a more favorable scaling than CASCI or CASSCF. Further work shows that DMRG-TCCSD performs well for difficult multireference systems and also reports a version of the method employing local pair natural orbital.\cite{Faulstich2019,Antalk2019} The authors suggest that the inclusion of high-order external amplitudes, as done in the RMR CC method, could lead to even more accurate results.\cite{Veis2016}

In this research, we reexamine the simplest type of size-extensive externally corrected coupled cluster, referred here as CASCI-CC. In this approach, the external correction comes from a small FCI computation within an active space. A shortcoming of this strategy is the prohibitive cost for large active spaces. Therefore, we investigate the substitution of CASCI for the adaptive configuration interaction (ACI) method developed by Schriber and Evangelista.\cite{Schriber2016,Schriber2017} ACI is a modern selected CI algorithm that is suitable for extended active spaces. In his review of externally and internally corrected coupled cluster methods, Paldus\cite{Paldus2016} highlights desirable properties for the external source of high-order clusters. These properties are: (i) universal availability: the ACI method is a general approach that relies on known CI technology, thus it can be extended for different systems; (ii) size consistent: as an approximation to FCI, the ACI method can numerically converge to a size consistent answer; (iii) account for non dynamic correlation: ACI selects the most important determinants to build the wave function, including higher-than-pair excitations, thus accounting for nondynamical correlation; (iv) be systematically improvable: the strongest suit of the ACI method is its dependency on a pre selected $\sigma$ parameter that approximately controls the precision of the method (see Equation \ref{sigma}). 
\begin{align}
|E_\text{FCI} - E_\text{ACI}| \approx \sigma \label{sigma}
\end{align}
In this light, one can tune the ACI method arbitrarily towards the FCI limit. The ACI externally corrected CCSD is henceforth denoted as ACI($\sigma$)-CCSD and, when perturbative triples are included, ACI($\sigma$)-CCSD(T), where $\sigma$ is on the energy scale of \Eh{}. For example, ACI(10) denotes a $\sigma$ of 10 \Eh{}.

\section{Theory}
\subsection{Adaptive Configuration Interaction}

\begin{figure}
  \centering
  \includegraphics[width=0.7\textwidth]{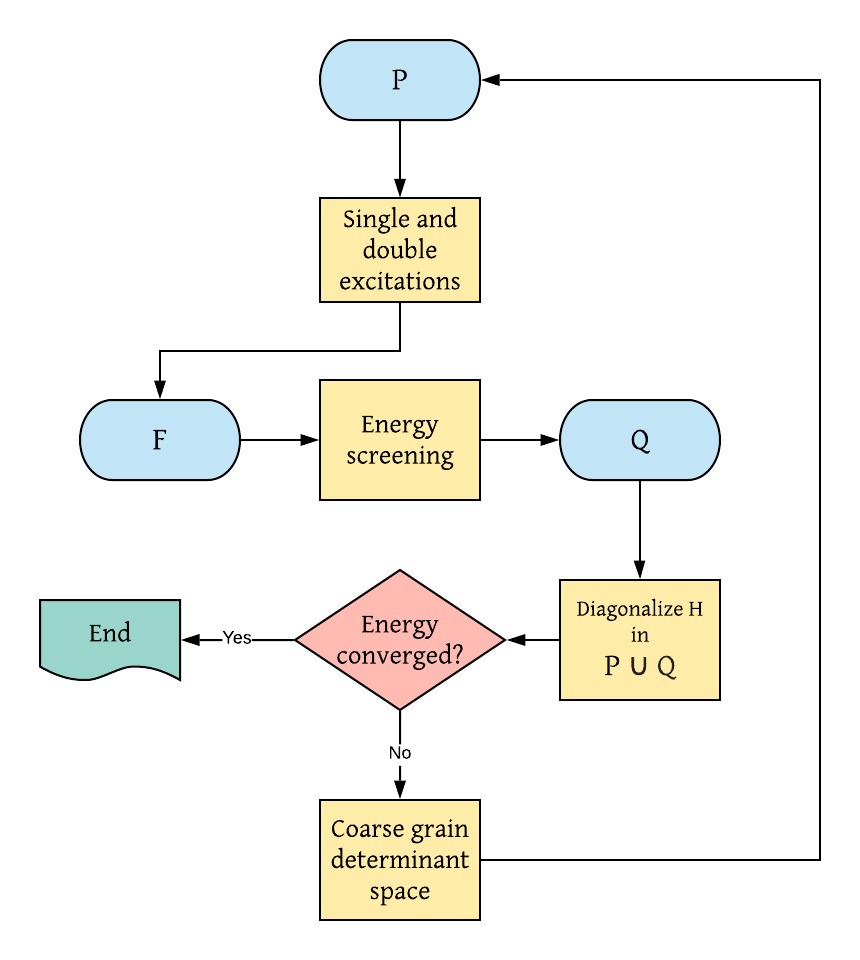}
  \caption{Summary of the ACI procedure. Determinants in the set $F$ are screened using Equation \ref{eqscreen}. Coarse-graining is applied until Equation \ref{eqcoarsegrain} is satisfied.}
  \label{fig:aciflowchart}
\end{figure}

The following is a brief description of the ACI algorithm; a more thorough discussion of the method can be found in the previous literature.\cite{Schriber2016, Schriber2017} We start with set of determinants $P$, which has an associated normalized wave function $\ket{\Psi_P} = \sum c_p \ket{\Phi_p}$ for $p \in P$, where $c_p$ are the CI expansion coefficients. A new set, $F$, is formed by taking single and double excitations of $P$. For each $\ket{\Phi_f}$, where $f \in  F$, we associate an energy contribution $\epsilon$ determined by
\begin{align}
    \epsilon_f &= \frac{\Delta}{2} - \sqrt{\frac{\Delta^2}{4}+V^2} \label{eqscreen}\\[2mm]
    \Delta = E_f - E_P &= \bra{\Phi_f}H\ket{\Phi_f} - \bra{\Psi_P}H\ket{\Psi_P} \\[2mm]
    V = \bra{\Phi_f}H\ket{\Psi_P} &= \sum_k c_k \bra{\Phi_f}H\ket{\Phi_k} \;\; k\in P
\end{align}
The determinants with smallest $|\epsilon|$ are removed from $F$ until the sum of their contributions reaches a predefined parameter $\sigma$. After this removal, the remaining determinants form the space $Q$. The updated model space ($M$) is created as $M = P \cup Q$. For this new space, the Hamiltonian matrix is constructed and diagonalized. These steps are repeated until convergence is reached. However, before each new iteration, the set $M$ undergoes a coarse graining procedure, where the determinants with the lowest contribution to the wave function ($c^2$) are removed until 
\begin{align}
    \sum_m c_m^* c_m < 1-\gamma \sigma \label{eqcoarsegrain}
\end{align}
where $\gamma$ is set to 1 $E_\mathrm{h}^{-1}$ as recommended by the authors.\cite{Schriber2017} Figure \ref{fig:aciflowchart} schematically summarizes the ACI procedure. 

\subsection{Externally Corrected Coupled-Cluster}
In CC theory, the energy of the system is determined utilizing only the $T_1$ and $T_2$ amplitudes. However, despite higher-order amplitudes not being present in the energy expression, they are coupled with $T_1$ and $T_2$ through the full CC equations. In looking at the amplitude equations from the CCSDTQ level of theory, it is readily seen that $T_3$ and $T_4$ amplitudes interact directly with $T_1$ and $T_2$. Defining the orbital-energy (diagonal Fock matrix elements) differences $D_i^a = f_{ii} - f_{aa}$ and $D_{ij}^{ab} = f_{ii} + f_{jj} - f_{aa} - f_{bb}$, higher-order contributions to $T_1$ and $T_2$ are
\begin{align}
    t_{i}^{a}D_{i}^{a} &\leftarrow \bra{\Phi_i^a}H_N\T_3 \rf_c \label{ect1}\\
    t_{ij}^{ab}D_{ij}^{ab} &\leftarrow \bra{\Phi_{ij}^{ab}}H_N(\T_3+\T_4+\T_1\T_3) \rf_c \label{ect2}
\end{align}
The traditional CCSD approach can be viewed as an approximation to the full CC where $T_3 = T_4 = 0$. As discussed in the literature,\cite{Paldus2016} any reasonable approximation for those amplitudes would yield better results than simply setting them to zero. Therefore, if we can estimate the contributions of $\T_3$ and $\T_4$ onto $\T_1$ and $\T_2$, we can recover some nondynamical correlation energy. In this work, the ACI method is utilized as the external method to approximate the the $T_3$ and $T_4$ amplitudes
\begin{align}
    \T_3 \approx \T_3^\text{ACI} \\
    \T_4 \approx \T_4^\text{ACI}
\end{align}
where $\T^\text{ACI}$ is obtained from a cluster analysis of the ACI wave function. For example, the equation relating $c_{ij}^{ab}$ to cluster amplitudes is obtained as
\begin{align}
c_{ij}^{ab} &= \bra{\Phi_{ij}^{ab}}e^{\T}\ket{\Phi_0} \\
&= \bra{\Phi_{ij}^{ab}}\T_2 + \frac{1}{2}T_1^2\ket{\Phi_0} \\
&= t_{ij}^{ab} + t_i^a t_j^b - t_j^a t_i^b
\end{align}
Recursive equations for determining $t_i^a$, $t_{ij}^{ab}$, $t_{ijk}^{abc}$, and $t_{ijkl}^{abcd}$ can be found in the literature. \cite{Lehtola2017} Once higher-order amplitudes are available, external corrections for $T_1$ and $T_2$ amplitudes are computed. Equations \ref{ect1} and \ref{ect2} produce known algebraic expressions from the CCSDTQ method, for completeness we repeat them here 
\begin{align}
t_i^a D_i^a &\leftarrow \frac{1}{4}t^{aef}_{imn} v^{mn}_{ef}  \label{ect1eq}\\[2mm]
t_{ij}^{ab} D_{ij}^{ab} &\leftarrow f^{m}_{e} t^{abe}_{ijm} +t^{e}_{m} t^{abf}_{ijn} v^{mn}_{ef} +\frac{1}{2} \hat{P}(ab)[t^{aef}_{ijm} v^{bm}_{ef}] -\frac{1}{2} \hat{P}(ij)[t^{abe}_{imn} v^{mn}_{je}] \label{ect2eq} \nonumber\\
\phantom{t_{ij}^{ab} D_{ij}^{ab}}&\phantom{\leftarrow} +\frac{1}{2} \hat{P}(ij) [t^{e}_{i} t^{abf}_{jmn} v^{mn}_{ef} ] +\frac{1}{2}\hat{P}(ab)[t^{a}_{m} t^{bef}_{ijn} v^{mn}_{ef}] 
 + \frac{1}{4}t_{ijmn}^{abef}v^{mn}_{ef} 
\end{align}
where the summation over repeated indices on the right hand side is implied and $\hat{P}(pq)$ is the antisymmetric permutation operator. We reinforce that in ecCC, $T_3$ and $T_4$ amplitudes are known \textit{a priori}, these terms are obtained once from the ACI wave function and their contributions (Equations \ref{ect1eq} and \ref{ect2eq}) are fixed throughout the computation. In principle, the $T_1 T_3$ terms can be updated for each iteration, as the $t_i^a$ amplitudes change. However, as typically done in ecCC, we compute this term one time using the $T_1^\text{ACI}$ and neglect its changes. The difference between the two strategies seems to be insignificant.\cite{Li1997}

\subsection{Perturbative Triples Correction (T)} 

The inclusion of perturbative energy corrections has been done in ecCC by treating connected triples outside the external correction in the same way it is done in the CCSD(T) method.\cite{Li2006} In our implementation, the perturbative correction for triple amplitudes is included using a standard \textit{ijk} algorithm.\cite{Rendell1991} However, the energy correction associated with each amplitude $t_{ijk}^{abc}$ is included only if the corresponding determinant is not part of the ACI wave function. 
\begin{align}
    E_{(T)} \leftarrow t_{ijk}^{abc}\;\; \text{if} \;\; \ket{\Phi_{ijk}^{abc}} \notin \; \text{ACI}
\end{align}

\section{Computational Details}

The ACI, CC and ecCC methods were implemented using our under development, open-source, electronic structure code \textsc{Fermi},\cite{fermi} written in the programming language \textsc{Julia}.\cite{julia} CCSDT and CCSDTQ computations were performed using the MRCC package.\cite{mrcc1,mrcc2,mrcc3} Potential energy curves were interpolated and vibrational energy levels were obtained using the matrix Numerov method.\cite{numerov} Dunning's correlation consistent cc-pVXZ basis set were employed for all computations.\cite{Dunning1989}
Unless otherwise specified, restricted canonical Hartree--Fock orbitals are utilized and Hartree--Fock determinant is taken as the reference for CC and ecCC computations.

\section{Results and Discussion}
\subsection{Beryllium Dimer}

\begin{figure}[H]
    \centering
    \includegraphics[scale=0.5]{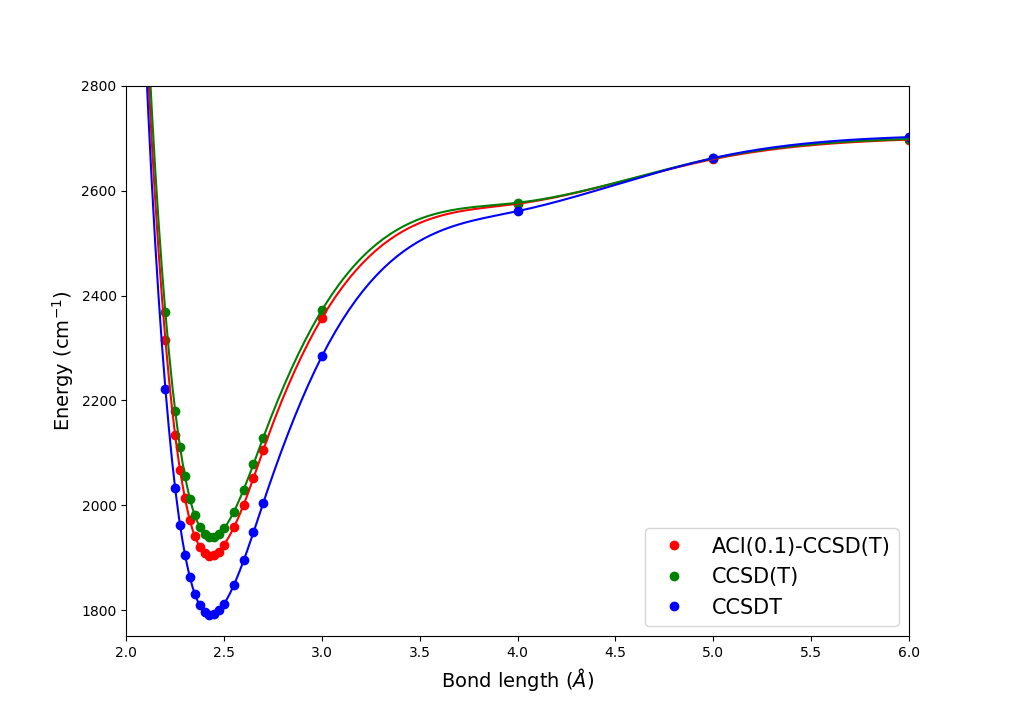}
    \caption{Potential energy curve for the beryllium dimer. Energies are shown with respect to the dissociation limit at the corresponding level of theory. All electrons are correlated. ACI computed with a (8\textit{e},10\textit{o}) active space.}
    \label{figure:pec}
\end{figure}

\ce{Be_2} is a challenging system for both theory and experiment. For a complete and recent discussion on this topic, we 
refer the reader to Lesiuk \textit{et al.}\cite{Lesiuk2019} and Merritt \textit{et al.}\cite{Merritt2009}. Past work has been concluded that the elucidation of the \ce{Be2} bond must include a description of strong correlation.\cite{ElKhatib2014} 
We examine the potential energy curve of the beryllium dimer using CC and ecCC methods. 
Computed low lying vibrational levels ($\nu$ = 0--4) using the cc-pVQZ basis set and the active space of eight electrons in ten orbitals (8\textit{e},10\textit{o}) are shown in Table \ref{table:bevib}, and corresponding selected potential energy curves are depicted in Figure \ref{figure:pec}. The CCSD(T) method successfully describes the fundamental transition, being only 1.5 \cm{} away from the experimental value. However, the quality of the results decays rapidly for the higher vibrational energy levels; already for $\nu_3$ the deviation is greater than 26 \cm{}. The ACI(0.1)-CCSD(T) result for the fundamental is 4 \cm{} away from the experimental value; however it improves upon CCSD(T) for higher energy levels; for $\nu_2$ and $\nu_3$ errors with respect to experiment are only 1 \cm{} and 5 \cm{}. The overall effect of post-CCSD(T) terms is to raise vibrational energy levels. In fact, the CCSDT method seems to perform poorly when compared to CCSD(T), as all energies predicted at this level are above the experimental results. Moreover, it can be seen in Figure \ref{figure:pec} that utilizing an adaptive method as the source of external correction has not introduced noticeable discontinuities in the potential energy curve. 
Table \ref{table:beDe} presents results for the \ce{Be2} well depth using the cc-pV5Z basis set. These values were computed as the difference between the energy of the dimer at the minimum (determined using the cc-pVQZ basis set, see Table \ref{table:bevib}) using the (8\textit{e},10\textit{o}) active space and two isolated beryllium atoms using (4\textit{e},5\textit{o}) as the active space. It can be seen that the external correction brings the CCSD(T) result down by 25 \cm{} outperforming the CCSDT method, which seems to overestimate the well depth. When compared to experiment CCSD(T) and ACI-CCSD(T) have similar errors, around 12 \cm{}. However, the expanded Morse oscillator model of the potential employed by Merritt \textit{et al.}\cite{Merritt2009} to determine $D_e$ has been criticised by Lesiuk \textit{et al.}\cite{Lesiuk2019} whose exhaustive theoretical work predicted a slightly deeper potential well. Similarly, many other theoretical studies found well depths greater than 935 \cm{}.\cite{Martin1999,Patkowski2007,Koput2011} Thus, we believe the shift in the energy due to the external correction can be seen as an improvement.
The ability of ACI to produce CASCI quality corrections is also illustrated in this example. While the $\sigma= 1$ \Eh{} case does not seem to contain all significant information, a tighter $\sigma = 0.1$ \Eh{} reproduces the full CASCI correction very well. We highlight that using the CASCI method, the active space size contains 44100 determinants (no symmetry considered). Whereas the ACI(0.1) optimized model space comprises less than 90 determinants when $R = 2.425$ \AA. More interesting, for the beryllium atoms the ACI wave function does not contain higher than doubly excited determinants, even when we tried a very tight ACI(0.01). This means that no external correction is added to the beryllium atoms, which seems problematic suggesting the atoms are not being treated on the same footing as the molecule. However, a comparison with CASCI results, which explicitly include all determinants within the active space, confirms that, when computing $D_e$, the external correction for the \ce{Be} atom is negligible.

\begin{table}
    \setlength{\tabcolsep}{5pt}
    \caption{Beryllium dimer bond length ($R_e$, in angstrom) and low lying vibrational frequencies ($\Delta \nu = \nu_i - \nu_0$ in \cm{} for i = 1--4)  for different methods. All electrons are correlated and the active space used included eight electrons and the ten lowest orbitals (8\textit{e},10\textit{o}). The cc-pVQZ basis set was employed in all cases. Values in parenthesis are errors with respect to experimental values.}
    \begin{tabular}{c c c c c c}
    \hline
    Method & $R_e$/\AA & $\nu_1$ & $\nu_2$ & $\nu_3$ & $\nu_4$ \\[2mm]
    \hline
    CCSD(T)                 & 2.436  & 221.1 (\textit{1.5})  & 386.3  (\textit{10.8})  & 492.4  (\textit{25.7})  & 535.3  (\textit{59.5})  \\[2mm]
    ACI(1)-CCSD(T)    & 2.432  & 224.5  (\textit{1.9})  & 393.5  (\textit{3.6}) & 505.5  (\textit{12.6}) & 553.4  (\textit{41.4})  \\[2mm]
    ACI(0.1)-CCSD(T) & 2.430  & 226.6 (\textit{4}) & 398.1 (\textit{1}) & 512.9 (\textit{5.2}) & 563.2 (\textit{31.6})  \\[2mm]
    CASCI-CCSD(T) & 2.428  & 228.0 (\textit{5.4}) & 400.22 (\textit{3.1}) & 515.3 (\textit{2.8}) & 566.2 (\textit{28.6})  \\[2mm]
    CCSDT                    & 2.432  &  235.0 (\textit{12.4})  & 412.0 (\textit{14.9})  & 555.8 (\textit{37.7}) & 636.1 (\textit{41.3})  \\[2mm]
    ecCCSDt-CASSCF\textsuperscript{\emph{a}}  & 2.423  & 242.5 (\textit{19.9}) &  & & \\[2mm]
    \hline
    Exp.\textsuperscript{\emph{b}}            & 2.454  & 222.6  & 397.1  & 518.1 & 594.8 \\
    \hline 
    \textsuperscript{\emph{a}} Xu and Li\cite{Xu2015};
    \textsuperscript{\emph{b}} Merritt \textit{et al.} \cite{Merritt2009}
    \end{tabular}
    \label{table:bevib}
\end{table}

\begin{table}
    \setlength{\tabcolsep}{5pt}
    \centering
    \caption{Well depth ($D_e$) for the beryllium dimer. All electrons are correlated and the active space included four electrons and the ten lowest orbitals (8\textit{e},10\textit{o}). The cc-pV5Z basis set was employed in all cases. Geometries were determined at the cc-pVQZ level of theory.}
    \begin{tabular}{c c}
    \hline
    Method & $D_e$/$cm^{-1}$ \\
    \hline
    CCSD(T)          & 919   \\[2mm]
    CCSDT            & 1082  \\[2mm]
    ACI(1)-CCSD(T)   & 933   \\[2mm]
    ACI(0.1)-CCSD(T) & 943   \\[2mm]
    CASCI-CCSD(T)    & 943   \\[2mm]
    Reference Value (Theory)\textsuperscript{\emph{a}}              & 934.6 $\pm$ 2.5 \\
    Reference Value (Exp.)\textsuperscript{\emph{b}}             & 929.7 $\pm$ 2.5 \\
    \hline
    \textsuperscript{\emph{a}} Lesiuk \textit{et al.}\cite{Lesiuk2019}; 
    \textsuperscript{\emph{b}} Merrritt \textit{et al.}\cite{Merritt2009}.
    \end{tabular}
    \label{table:beDe}
\end{table}

\subsection{Carbonyl Oxide Heat of Formation}

Carbonyl oxide is the simplest Criegee intermediate, it has received great attention due to the role it might play in atmospheric chemistry.\cite{Khan2018} Since this molecule possesses mild multireference character, many composite approaches applied to these systems included additive corrections such as $E_\text{CCSDT(Q)} - E_\text{CCSD(T)}$ with a relatively small basis set.\cite{Long2016,Misiewicz2018}

\begin{figure}[H]
    \centering
    \includegraphics[width=\textwidth]{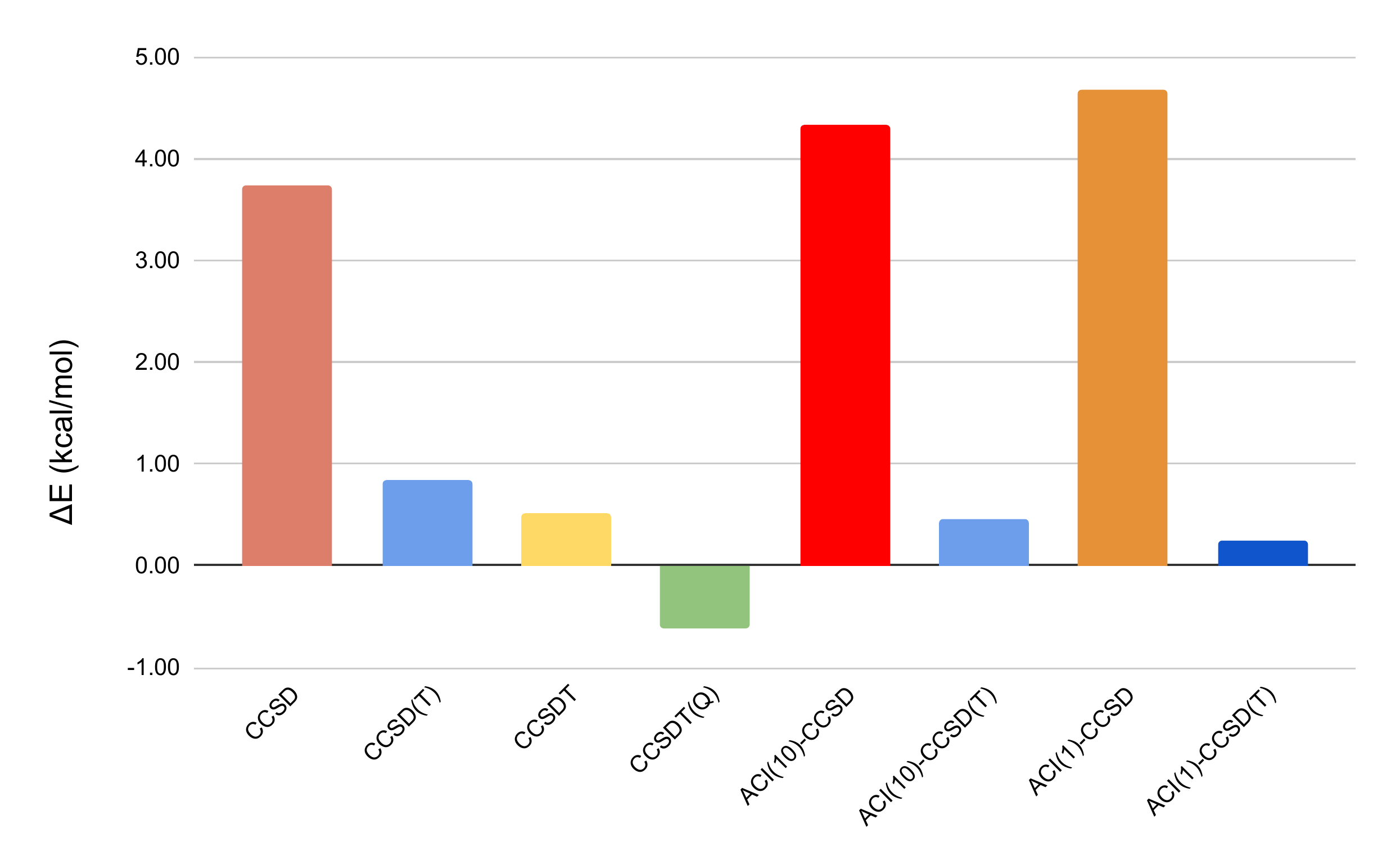}
    \caption{Errors in the \ce{H_2 + CO_2 -> H_2COO} reaction energy compared to a CCSDTQ computation in \kcal{} using the cc-pVDZ basis set. ACI utilized a full valence active space. Core electrons were frozen for all methods. Geometries optimized at the CCSD(T)/ANO2 level of theory.}
    \label{figure:criegee}
\end{figure}

To study the role of post-CCSD(T) contributions on the carbonyl oxide energy profile, we examine the following elementary reaction
\begin{equation} \tag{R1}
    \ce{H2(g) + CO2(g) -> H2COO(g)}
\end{equation}
Reliable data for the enthalpy of formation of \ce{CO2} is available;\cite{Ruscic2005} hence this reaction gives information that leads to the heat of formation of the carbonyl oxide. Figure \ref{figure:criegee} depicts errors in the electronic energies for reaction R1 employing the cc-pVDZ basis set with respect to the CCSDTQ result. At the CCSD level the error with respect to the CCSDTQ result is 3.7 \kcal; this number is greatly improved with the inclusion of perturbative triple corrections; CCSD(T) is 0.84 \kcal{} away from the CCSDTQ result. If chemical accuracy is desired and assuming CCSDTQ is a good approximation for FCI, then CCSD(T) seems to suffice. The CCSDT(Q) is the only method to present a negative deviation, which indicates that the perturbative quadruple correction may be favoring disproportionately the Criegee intermediate. This raises problems when CCSDT(Q) is used to assess post-CCSD(T) contributions. For this system, the full T plus (Q) correction to the CCSD(T) energy [$E_\text{CCSDT(Q)} - E_\text{CCSD(T)}$] is --1.45 \kcal{} which, contrary to the result discussed above, suggests that CCSD(T) is not within 1.0 \kcal{} of accuracy. 

The ACI computation was performed with $\sigma = 10$ and $1$ \Eh{} utilizing a full valence active space, that is (18\textit{e},14\textit{o}) for \ce{H2COO} and (16\textit{e},12\textit{o}) for \ce{CO2}. The external correction pushed the CCSD results further away from the reference data. However, the inclusion of perturbative triples improves the results dramatically; the ACI(1)-CCSD(T) result is only 0.24 \kcal{} above the CCSDTQ value. The fact that the ACI-CCSD results are somewhat worse than the standard CCSD suggests some imbalance in the recovery of correlation energy. The correlation energy increases by 0.7\% for \ce{CH2OO} and 1.1\% for \ce{CO2} when external correction is included into CCSD.  The subsequent inclusion of perturbative triples further increases the correlation energies by 4.4\% for \ce{CH2OO} and 3.4\% for \ce{CO2}. Therefore, the external $T_3^\text{ACI}$ correction appears to be improving the \ce{CO_2} absolute energy more rapidly than that of the carbonyl oxide. The inclusions of all triples (some through external corrections and the remaining through the (T) strategy) gives the two molecules a more balanced description. It is worth reinforcing that the perturbative triples energy corrections in the ecCC framework is not the same as in the tradition coupled cluster. Since ecCC is an amplitude corrected method, the final $t_{i}^{a}$ and $t_{ij}^{ab}$ are different, and therefore the (T) energy will also be different. Moreover, perturbative corrections associated with $t_{ijk}^{abc}$ inside the ACI are not included to avoid double counting. It has been long known that CCSD needs to be augmented with corrections for triples in order to achieve high accuracy. This system provides an example that this could also be the case for externally corrected methods. Finally, ACI-CCSD(T) can be used successfully to assess higher-order contributions to the CCSD(T) energy. For reaction R1, $E_\text{ACI(1)-CCSD(T)} - E_\text{CCSD(T)}$ is --0.60 \kcal{}, which, in this case, is a more realistic assessment than the one from CCSDT(Q) at lower cost.

\subsection{Cyclobutadiene Automerization}
\begin{figure}[H]
    \centering
    \includegraphics[width=\textwidth]{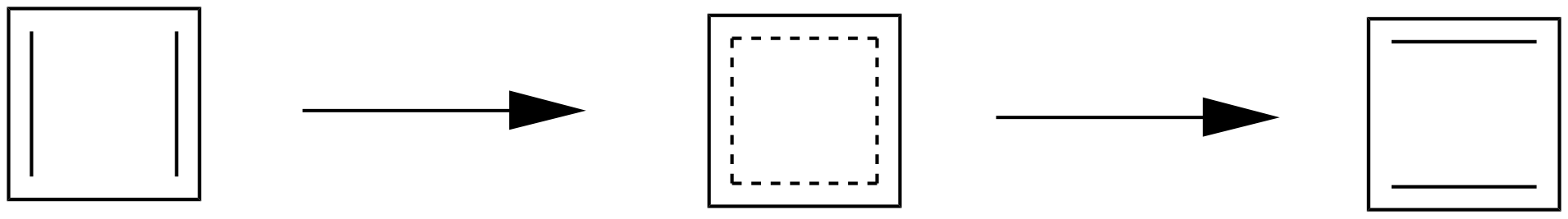}
    \caption{Automerization reaction of the cyclobutadiene}
    \label{figure:automerization}
\end{figure}

The automerization of cyclobutadiene, depicted in Figure \ref{figure:automerization}, is a classic example of a multireference transition state. The reaction starts at the rectangular $D_{2h}$ geometry reaching a square $D_{4h}$ transition state and finishes at a new rectangular geometry, degenerate with the initial one. The experimental value is estimated to lie within the wide range of 1.6--10 \kcal{}.\cite{Whitman1982} 

Table \ref{table:automerization} presents results for the automerization barrier height for a few selected methods including ACI(10)-CCSD(T). These values do not include zero-point vibrational energy corrections. For the ACI computations, we employed a full valence active space with 20 electrons in 20 orbitals (20\textit{e},20\textit{o}). While it is clear that the ACI(10) captured some important contributions, improving CCSD(T) results by more than 3 \kcal{}, the external correction is only half as much as the one produced by RMR CCSD(T).\cite{Li2009} The ACI wave function contains two dominant configurations, one of which is the Hartree--Fock determinant. To properly account for the correlation energy, one must include all single and double excitations from these determinants as it is done in MRCI. Thus, it is not surprising that the RMR CCSD(T), which uses MRCI as a source of external corrections, describes this system better. We believe an augmentation of the ACI wave function, allowing it to include important determinants outside the initial active space, would produce competitive results. This is similar to the proposed PSCI CCSD by Peris \textit{et al.}\cite{Peris1999} It is important to note that the ACI-CCSD(T) method is a simpler and less demanding approach than the RMR CCSD(T). Nevertheless, ACI-CCSD(T) can provide a natural starting point for more complex external corrections. 

\begin{table}
    \setlength{\tabcolsep}{5pt}
    \centering
    \caption{Selected MR CC results for the pure electronic barrier height of the cyclobutadiene automerization using the cc-pVDZ basis set. ACI utilized a full valence active space (20\textit{e},20\textit{o})}
    \begin{tabular}{c c c}
    \hline
    Method & Barrier height/\kcal{}   & Reference\\
    \hline
    CCSD(T)         & 16.2  & This work \\[2mm]
    ACI(10)-CCSD(T) & 12.7  & This work \\[2mm]
    RMR-CCSD(T)     & 7.2   & \citenum{Li2009}  \\[2mm]
    RMR-CCSD(T)\textsuperscript{\emph{a}}     & 5.9   & \citenum{Li2009}  \\[2mm]
    MR-AQCCSD       & 7.3   &   \citenum{EckertMaksi2006} \\
    \hline
    \textsuperscript{\emph{a}} MCSCF orbitals.
    \end{tabular}
    \label{table:automerization}
\end{table}

\section{Conclusions}
We introduce in this work the ACI($\sigma$)-CCSD(T) method. It may be seen as an approximation to the simplest size-consistent externally corrected method, using a small FCI within an active space as the source of external corrections. However, employing the ACI allows us to compute corrections with extended active spaces. We anticipate that a production level code could be used on actives spaces much larger than the (20\textit{e},20\textit{o}) utilized here. 
In the three systems examined the ACI($\sigma$)-CCSD(T) performed satisfactorily, particularly well for \ce{Be_2} and \ce{CH_2OO}, where the results were comparable or superior to the CCSDT method.

Despite the fact that the ACI($\sigma$)-CCSD(T) method improved upon CCSD(T) results in all cases studied, the computed cyclobutadiene automerization energy fell short when compared to more involved methods, such as RMR CCSD (which is itself an ecCC method). This highlights the limitation of using a restricted active space to compute external corrections. Future work should explore a way to screen determinants without the limitations of an active space, keeping the computational cost manageable. Furthermore, using ACI as the FCI solver, orbital optimization can also considered. In future work, we hope to extend the ACI-CCSD(T) method to open-shell systems.

\begin{acknowledgement}
The authors thank the support from the National Science Foundation, Grant No. CHE-1661604.

\end{acknowledgement}

\begin{suppinfo}

Raw electronic energies and Cartesian coordinates of all molecules.

\end{suppinfo}

\bibliography{main.bib}

%% graphical table of contents.
%%%%%%%%%%%%%%%%%%%%%%%%%%%%%%%%%%%%%%%%%%%%%%%%%%%%%%%%%%%%%%%%%%%%%

\begin{tocentry}

\begin{figure}[H]
    \centering
    \includegraphics[width=8.25cm]{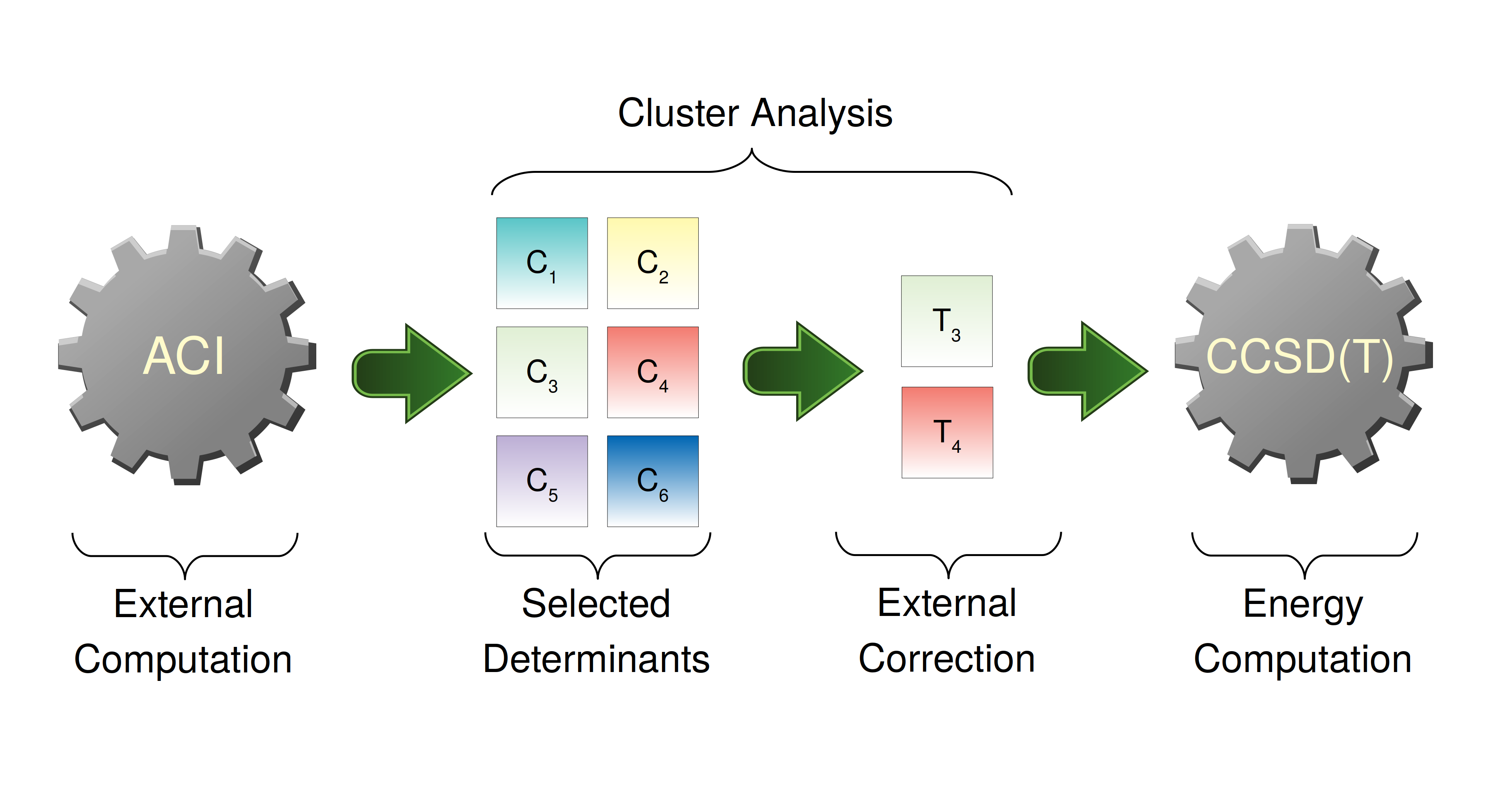}
    \caption{For Table of Contents Only}
\end{figure}
\end{tocentry}

\end{document}